\documentclass[referee]{aa} 
\usepackage{graphicx}
\usepackage{slashbox}
\usepackage{txfonts}
\begin{document}

\title{Plasma environment effects on K lines of astrophysical interest}

\subtitle{II. Ionization potentials, K thresholds, radiative rates and Auger widths in Ne- through He-like iron ions (\ion{Fe}{xvii} -- \ion{Fe}{xxv})}

\author{J. Deprince
          \inst{1},
          M.A. Bautista
          \inst{2},
          S. Fritzsche
          \inst{3,4},
          J.A. Garc\'ia
          \inst{5,6},
          T. Kallman
          \inst{7},
          C. Mendoza
          \inst{2},
          P. Palmeri
          \inst{1},
          \and
          P. Quinet
          \inst{1,8}
             }
\institute{Physique Atomique et Astrophysique, Universit\'e de Mons -- UMONS, B-7000 Mons, Belgium
         \and
             Department of Physics, Western Michigan University, Kalamazoo, MI 49008, USA
         \and
             Helmholtz Institut Jena, 07743 Jena, Germany
         \and
             Theoretisch Physikalisches Institut, Friedrich Schiller Universit\"at Jena, 07743 Jena, Germany
         \and
             Cahill Center for Astronomy and Astrophysics, California Institute of Technology, Pasadena, CA 91125, USA
         \and
             Dr. Karl Remeis-Observatory and Erlangen Centre for Astroparticle Physics, Sternwartstr.~7, 96049 Bamberg, Germany
         \and
             NASA Goddard Space Flight Center, Code 662, Greenbelt, MD, USA
         \and
             IPNAS, Universit\'e de Li\`ege, Sart Tilman, B-4000 Li\`ege, Belgium\\
             \email{Pascal.Quinet@umons.ac.be}
             }
\titlerunning{Plasma effects on K lines in He- through Ne-like iron ions}

\authorrunning{Deprince et al.}

\date{Received ??; accepted ??}

\abstract
   {}
   {In the context of accretion disks around black holes, we estimate plasma-environment effects on the atomic parameters associated with the decay of K-vacancy states in highly charged iron ions, namely \ion{Fe}{xvii} -- \ion{Fe}{xxv}.}
   {Within the relativistic multiconfiguration Dirac--Fock (MCDF) framework, the electron--nucleus and electron--electron plasma screenings are approximated with a time-averaged Debye-H\"uckel potential.}
   {Modified ionization potentials, K thresholds, wavelengths, radiative emission rates and Auger widths are reported for astrophysical plasmas characterized by electron temperatures and densities respectively in the ranges $10^5{-}10^7$~K and $10^{18}{-}10^{22}$~cm$^{-3}$.}
   {We conclude that the high-resolution micro-calorimeters onboard
   future X-ray missions such as {\it XRISM} and {\it ATHENA} are expected to be sensitive to the lowering of the iron K edge due to the extreme plasma conditions occurring in accretion disks around compact objects.}

\keywords{Black hole physics -- Plasmas -- Atomic data -- X-rays: general}

\maketitle
%

\section{Introduction}

Accurate descriptions of K-shell atomic processes are essential in the analysis of the X-ray spectra from space telescopes such as {\it Chandra}, {\it XMM-Newton}, {\it Suzaku}, and {\it NuSTAR}. In this respect, we have been involved for the past two decades in  extensive computations of the relevant atomic data for cosmic abundant elements with atomic number $Z\leq 30$, iron in particular, but also trace elements with odd $Z$ \citep[see, for example,][]{pal02, pal03a, pal03b, pal08, pal11, pal12, pal16, bau03, bau04, kal04, gar05, gar09, gar11, men04, men17, men18, wit09, wit11, gor13, has14}.

This data collection has become a reference source in several atomic databases for astrophysical applications, e.g. \citet{bau01}, AtomDB\footnote{\url{http://www.atomdb.org}}, {\sc chianti} \citep{lan12}, and uaDB\footnote{\url{http://heasarc.gsfc.nasa.gov/uadb/}} as well as in spectral modeling codes such as {\sc xstar} \citep{kal01}, {\sc cloudy} \citep{fer13}, {\sc spex} \citep{kaa96}, and {\sc ion} \citep{net96}. However, these atomic parameters have been computed neglecting plasma-embedding effects, and consequently, they are not expected to be applicable at relatively high electron densities, say $n_e>10^{18}$~cm$^{-3}$ \citep{smi14}.

Absorption and emission of high-energy photons in dense plasmas occur in a wide range of astrophysical phenomena, among the most exciting perhaps are the X-rays produced and reprocessed in the regions close to compact objects such as black holes and neutron stars. In these systems the gas inflow caused by the strong gravitational potential forms an accretion disk from which copious X-ray radiation is emitted \citep[e.g.][]{kro99, don07}. Observed spectra are usually imprinted with atomic features, both in emission and absorption, from a wide variety of ionic species, the modeling of which provides reliable insights of the composition, temperature, and degree of ionization of the plasma \citep{ros05, gar10}. In the case of a black hole, its angular momentum can be inferred by modeling the distortion of the Fe K emission complex caused by the strong relativistic effects \citep[e.g.][]{rey13, gar14}.

X-ray emission lines from accreting sources, in particular the K$\alpha$ and K$\beta$ lines from Fe ions, are characterized by observed widths and shifts mostly implying an origin close to the compact object \citep{rey03}; that is, the innermost stable circular orbit in the case of a black hole, or the stellar surface in the case of a neutron star. Line intensities may thus provide estimates of key properties of these exotic dense-plasma environments, including the effects of both special and general relativity in the emitting region, that are not attainable through other observational windows. In order to analyze the K$\alpha$ and K$\beta$ emission lines to derive, for instance, the Fe ionic fractions and abundance, it is of paramount importance to rely on accurate radiative and Auger data to infer the emission and absorption rates under various extreme conditions.

Dynamical models of black-hole accretion flows appear to support densities as high as $10^{21}{-}10^{22}$~cm$^{-3}$ \citep{rei13, sch13,tom18,jia19}, but thus far, their effect on line emission has not been studied in detail. It is worth noting that iron-ion survival near a black hole requires such high densities to counteract the strong ionization, and therefore, plasma embedding effects on the ionic structure, K photoexcitation and photoionization, and both radiative and Auger decay become unavoidable issues.

Following previous work on the oxygen isonuclear sequence by \citet{dep19a}, we have estimated density effects on the atomic parameters associated with the K-vacancy states of highly charged iron ions. Multiconfiguration Dirac--Fock computations have been carried out for these species representing the plasma electron--nucleus and electron--electron shieldings with a time-averaged Debye-H\"uckel potential. \citet{dep19a} have shown that both of these plasma interactions must be taken into account. We use a combination of the GRASP92 code \citep{par96} to obtain the wave functions and RATIP \citep{fri12} to calculate the atomic parameters. We report a first set of results on the ionization potentials, K thresholds, transition wavelengths, radiative emission rates, and Auger widths for nine ions: from \ion{Fe}{xvii} (Ne-like) to \ion{Fe}{xxv} (He-like).

\section{Theoretical model}

\subsection{Relativistic Multiconfiguration Dirac--Fock Method}

Wave functions for the ionic species \ion{Fe}{xvii} -- \ion{Fe}{xxv} are obtained using the fully relativistic multiconfiguration Dirac--Fock (MCDF) method, implemented in the GRASP92 version \citep{par96} of the General-purpose Relativistic Atomic Structure Program (GRASP) initially developed by \citet{gra80}, \citet{mck80}, and \citet{gra88}. In this approach the atomic state functions (ASF) $\Psi$($\gamma$$J$$M_J$) are expanded in linear combinations of the configuration state functions (CSF) $\Phi$($\alpha$$_i$$J$$M_J$)
\begin{equation}
\Psi(\gamma J M_J) = \sum_i c_i \Phi (\alpha_i J M_J)\ .
\end{equation}
The CSF are in turn taken as linear combinations of Slater determinants constructed from monoelectronic spin-orbitals of the form
\begin{equation}
\varphi_{n \kappa m} (r, \theta, \phi) = \frac{1}{r} \left( \begin{array}{c} P_{n \kappa}(r) \chi_{\kappa m} (\theta,\phi) \\ i Q_{n \kappa}(r) \chi_{- \kappa m} (\theta,\phi) \end{array} \right)\ ,
\end{equation}
where $P_{n \kappa}$($r$) and $Q_{n \kappa}$($r$) are, respectively, the large and small components of the radial wave function, and the angular functions $\chi$$_{\kappa m}$($\theta$,$\phi$) are spinor spherical harmonics. The $\alpha$$_i$ coefficients represent all the one-electron and intermediate quantum numbers needed to completely define the CSF, and $\gamma$ is usually chosen as the $\alpha$$_i$ corresponding to the CSF with the largest weight $|$$c_i$$|$$^2$. The $\kappa$ quantum number is given by
\begin{equation}\label{kappa}
\kappa = \pm (j + 1/2)\ ,
\end{equation}
$j$ being the electron total angular momentum. The sign before the parenthesis in Eq.~\eqref{kappa} corresponds to the coupling relation between the electron orbital momentum $l$ and its spin
\begin{equation}
l = j \pm 1/2 \ .
\end{equation}

The angular functions $\chi$$_{\kappa m}$($\theta$,$\phi$) are spinor spherical harmonics in the $lsj$ coupling scheme. Following \citet{gra07} their expression is given by
\begin{equation}
\chi_{\kappa m} (\theta,\phi) = \sum_{\sigma = \pm 1/2} \langle l, m-\sigma, \frac{1}{2}, \sigma | l, \frac{1}{2}, m \rangle Y_{l}^{m-\sigma}(\theta , \phi) \xi _{\sigma}
\end{equation}
with the two spinors
\begin{equation}
\xi _{\frac{1}{2}} = \left( {1 \atop 0} \right) \quad\quad \hbox{\rm and}\quad\quad \xi _{-\frac{1}{2}} = \left( {0 \atop 1} \right)\ .
\end{equation}
The radial functions $P$$_{n \kappa}$($r$) and $Q$$_{n \kappa}$($r$) are numerically represented on a logarithmic grid, and are required to be orthonormal within each $\kappa$ symmetry. In the MCDF variational procedure, the radial functions and expansion coefficients $c_i$ are optimized self-consistently.

In the present work the restricted active space (RAS) method is used to obtain the MCDF expansions for each ionic system. In this method electrons are excited from prescribed reference configurations to a given active set of orbitals; that is, the RAS is built up by considering all the single and double excitations from the ground and spectroscopic configurations to the $n=3$ configuration space. The list of reference configurations for each ionic system is specified as follows:

\begin{description}
  \item[\ion{Fe}{xvii}:] 1s$^2$2s$^2$2p$^6$, 1s$^2$2s$^2$2p$^5$3s
  \item[\ion{Fe}{xviii}:] 1s$^2$2s$^2$2p$^5$, 1s$^2$2s$^2$2p$^4$3s
  \item[\ion{Fe}{xix}:] 1s$^2$2s$^2$2p$^4$, 1s$^2$2s2p$^5$, 1s$^2$2p$^6$, 1s2s$^2$2p$^5$, 1s2s2p$^6$
  \item[\ion{Fe}{xx}:] 1s$^2$2s$^2$2p$^3$, 1s$^2$2s2p$^4$, 1s$^2$2p$^5$, 1s2s$^2$2p$^4$, 1s2s2p$^5$, 1s2p$^6$
  \item[\ion{Fe}{xxi}:] 1s$^2$2s$^2$2p$^2$, 1s$^2$2s2p$^3$, 1s$^2$2p$^4$, 1s2s$^2$2p$^3$, 1s2s2p$^4$, 1s2p$^5$
  \item[\ion{Fe}{xxii}:] 1s$^2$2s$^2$2p, 1s$^2$2s2p$^2$, 1s$^2$2p$^3$, 1s2s$^2$2p$^2$, 1s2s2p$^3$, 1s2p$^4$
  \item[\ion{Fe}{xxiii}:] 1s$^2$2s$^2$, 1s$^2$2s2p, 1s$^2$2p$^2$, 1s2s$^2$2p, 1s2s2p$^2$, 1s2p$^3$
  \item[\ion{Fe}{xxiv}:] (1s+2s+2p)$^3$
  \item[\ion{Fe}{xxv}:] 1s$^2$, 1s2s, 1s2p
\end{description}

The computations are carried out with the extended average level (EAL) option, optimizing a weighted trace of the Hamiltonian using level weights proportional to ($2J+1$), and they are completed with the inclusion of the relativistic two-body Breit interaction and the quantum electrodynamic corrections (QED) due to the self-energy and vacuum polarization. The MCDF ionic bound states generated by GRASP92 are then used in RATIP \citep{fri12} to compute the atomic structure and the radiative and Auger parameters associated with K-vacancy states.

\subsection{Plasma screening effects}

We use a Debye-H\"uckel (DH) potential to model the plasma screening effects on the atomic properties, which in atomic units (a.u.) is given by
\begin{equation}\label{dhpot}
V^{DH}(r,\mu)=-\sum_{i=1}^{N} \frac{Ze^{-\mu r_{i}}}{r_{i}} + \sum_{i>j}^{N} \frac{e^{-\mu r_{ij}}}{r_{ij}}\ .
\end{equation}
$N$ is the number of bound electrons, $r_i$ is the distance of the $i$th electron from the nucleus, and $r_{ij}$ is the distance between the $i$ and $j$ electrons. The plasma {\em screening parameter} $\mu$ is the inverse of the Debye shielding length $\lambda$$_{\rm De}$, and can be expressed in terms of the plasma electron density $n_e$ and temperature $T_e$ as
\begin{equation}
\mu = \frac{1}{\lambda_{\rm De}} = \left(\frac{4 \pi n_e}{k T_e}\right)^{1/2}\ .
\end{equation}

A given value of $\mu$ is then associated with a certain type of plasma environment. For example, according to the magnetohydrodynamic simulations reported by \citet{sch13} for accreting black holes with ten solar masses and an accretion rate of 10\%, the plasma conditions have been estimated at $T_e = 10^5{-}10^7$~K and $n_e = 10^{18}{-}10^{22}$~cm$^{-3}$. As shown in Table~\ref{Lambda}, this range corresponds to values of $\mu\lesssim 0.25$~a.u. for which the plasma-coupling parameter \citep{pie10}
\begin{equation}\label{ppar}
\Gamma = \frac{q^2}{4 \pi \epsilon _0 d k T}\ ,
\end{equation}
denoting the ratio of the electrostatic energy of neighboring particles to the thermal energy in the astrophysical plasma (predominantly protons and electrons), is always smaller (much smaller in most cases) than unity. This is in fact the validity condition for the statistically shielded DH potential to describe appropriately the screened Coulomb electrostatic interaction \citep[see, for example,][]{sah06, bel15, das16}. In Eq.~\eqref{ppar} $q$ is the charge of the neighboring particles separated by a typical distance $d~=~(3/4\pi n_e)^{1/3}$.

\begin{table}[h!]
  \caption{Screening parameter $\mu$ (a.u.) for different electron temperatures and densities. \label{Lambda}}
  \centering
  \small
  \begin{tabular}{c|ccccc}
  \hline\hline
  \noalign{\smallskip}
  \backslashbox{$T_e$\,(K)}{$n_e$\,(cm$^{-3}$)} & $10^{18}$ & $10^{19}$ & $10^{20}$ & $10^{21}$ & $10^{22}$ \\
  \noalign{\smallskip}
  \hline
  \noalign{\smallskip}
  $10^5$ & 0.002 & 0.008 & 0.024 & 0.077 & 0.242 \\
  $10^6$ & 0.001 & 0.002 & 0.008 & 0.024 & 0.077 \\
  $10^7$ & 0.000 & 0.001 & 0.002 & 0.008 & 0.024 \\
  \hline
  \end{tabular}
\end{table}

Therefore, when computing the atomic data with the RATIP code, we replace the electron--nucleus and electron--electron Coulomb interactions of the ionic system with the corresponding DH potential terms of Eq.~\eqref{dhpot} assuming values of the plasma screening parameter in the range $0\leq\mu\leq 0.25$~a.u. The applicability of this approach was recently discussed for the oxygen isonuclear sequence by \citet{dep19a}.

\begin{table}[b!]
  \caption{Computed ionization potentials (eV) for \ion{Fe}{xvii} -- \ion{Fe}{xxv} as a function of the plasma screening parameter $\mu$ (a.u.). NIST values are also given for comparison. \label{IP-plasma}}
  \centering
  \small
  \begin{tabular}{llccc}
  \hline\hline
  \noalign{\smallskip}
  Ion & NIST & $\mu = 0.0$ & $\mu = 0.1$ & $\mu = 0.25$	\\
  \hline
  \noalign{\smallskip}
  \ion{Fe}{xvii}  & $1262.7(7)$     & 1260.58 &	1214.75 & 1147.80 \\
  \ion{Fe}{xviii} & $1357.8(1.9)$	& 1357.09 &	1308.56	& 1237.68 \\
  \ion{Fe}{xix}   & $1460(3)$       & 1459.12 & 1407.92 & 1333.09 \\
  \ion{Fe}{xx}    & $1575.6(5)$     & 1573.48 & 1519.58 & 1440.80 \\
  \ion{Fe}{xxi}   & $1687(1)$       & 1689.13 & 1632.53 & 1549.75 \\
  \ion{Fe}{xxii}  & $1798.4(8)$     & 1797.82 & 1738.54 & 1651.80 \\
  \ion{Fe}{xxiii} & $1950.4(1.8)$	& 1950.49 & 1888.61 & 1798.43 \\
  \ion{Fe}{xxiv}  & $2045.759(7)$   & 2044.34 & 1979.79 & 1885.77 \\
  \ion{Fe}{xxv}   & $8828.1875(11)$ & 8836.74 & 8768.90 & 8667.86 \\
  \hline
  \end{tabular}
\end{table}

\section{Results and discussion}

\subsection{Ionization potentials and K-thresholds}

The ionization potentials (IP) and K thresholds we have computed with screening parameters in the range $0\leq\mu\leq 0.25$~a.u. are summarized in Tables~\ref{IP-plasma}--\ref{seuil-K-plasma}. For the isolated ions ($\mu$ = 0) we reproduce the IPs listed in the NIST spectroscopic database \citep{nist} to an accuracy better than 0.2\%. The inclusion of the DH potential leads to reductions of the IP and K-threshold energy positions that increase with $\mu$; namely 3\% and 8\% for $\mu=0.10$ and $\mu=0.25$, respectively, except for \ion{Fe}{xxv} for which the reduction is less than 2\% due to its large IP $>8600$~eV. For the same reason, the K-threshold reductions are only marginal ($<2\%$).

\begin{table}[h!]
  \caption{Computed K-thresholds (eV) for \ion{Fe}{xvii} -- \ion{Fe}{xxv} as a function of the plasma screening parameter $\mu$ (a.u.). \label{seuil-K-plasma}}
  \centering
  \small
  \begin{tabular}{lcccc}
  \hline\hline
  \noalign{\smallskip}
  Ion & $\mu = 0.0$	& $\mu = 0.1 $ & $\mu = 0.25$ \\
  \hline
  \noalign{\smallskip}
  \ion{Fe}{xvii}  &	7697.64	& 7651.45 & 7582.46	\\
  \ion{Fe}{xviii} & 7827.06 & 7778.17 & 7705.19 \\
  \ion{Fe}{xix}   & 7959.86 & 7908.27 & 7831.30 \\
  \ion{Fe}{xx}    & 8095.41 & 8041.11 & 7960.14 \\
  \ion{Fe}{xxi}   & 8246.57 & 8189.56 & 8104.56 \\
  \ion{Fe}{xxii}  & 8398.74 & 8339.03 & 8250.02	\\
  \ion{Fe}{xxiii} & 8558.75 & 8496.29 & 8403.31	\\
  \ion{Fe}{xxiv}  & 8689.49	& 8624.26 & 8527.21	\\
  \ion{Fe}{xxv}   & 8836.74	& 8768.90 & 8667.86	\\
  \hline
  \end{tabular}
\end{table}

\begin{figure}[h!]
  \centering
  \includegraphics[bb=60 160 484 571, clip=true, width=\columnwidth]{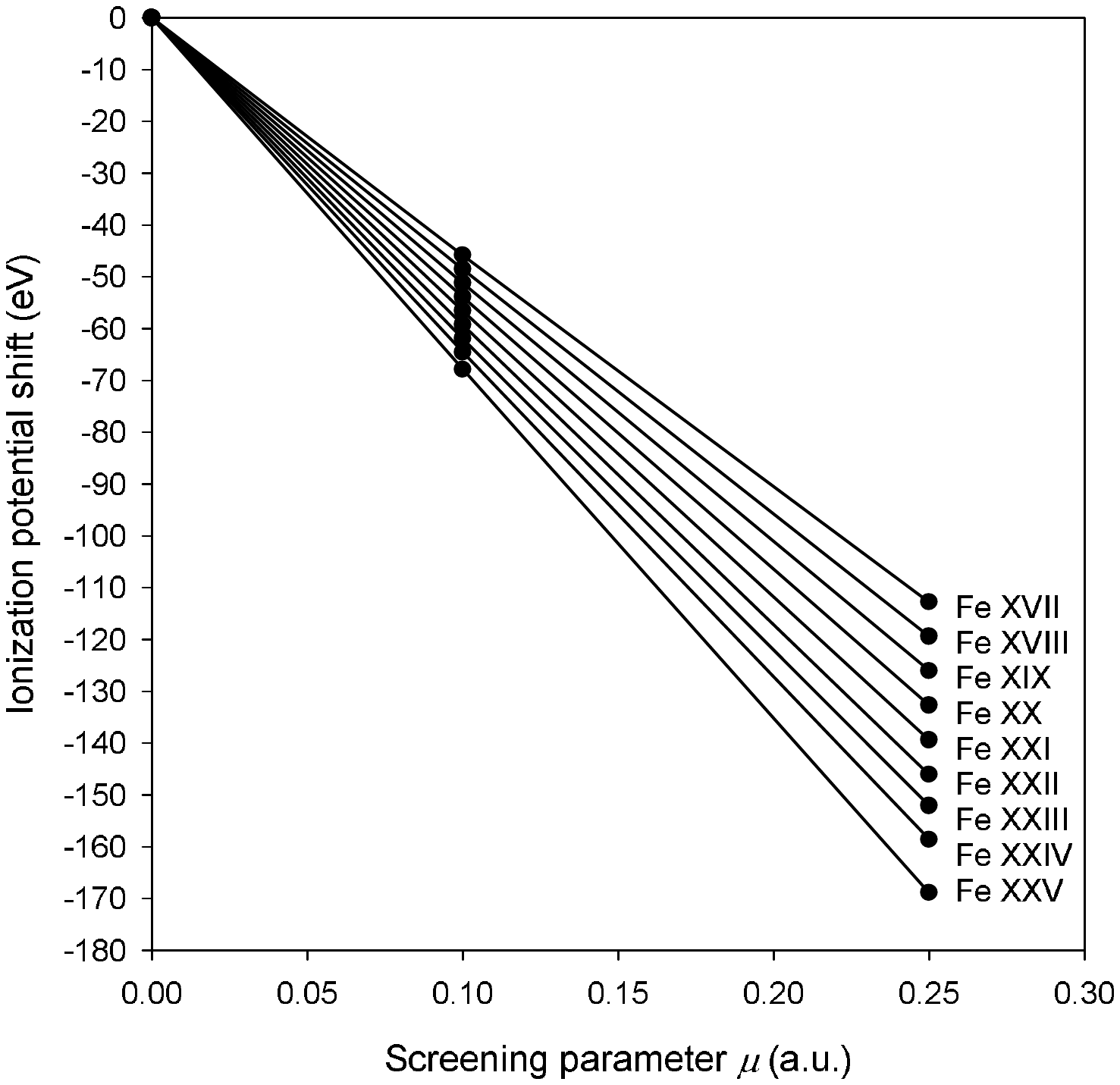}
  \caption{Ionization potential shifts in \ion{Fe}{xvii} -- \ion{Fe}{xxv} as a function of the plasma screening parameter $\mu$.} \label{Fig1}
\end{figure}

\begin{figure}[h!]
  \centering
  \includegraphics[bb=58 175 470 585, clip=true, width=\columnwidth]{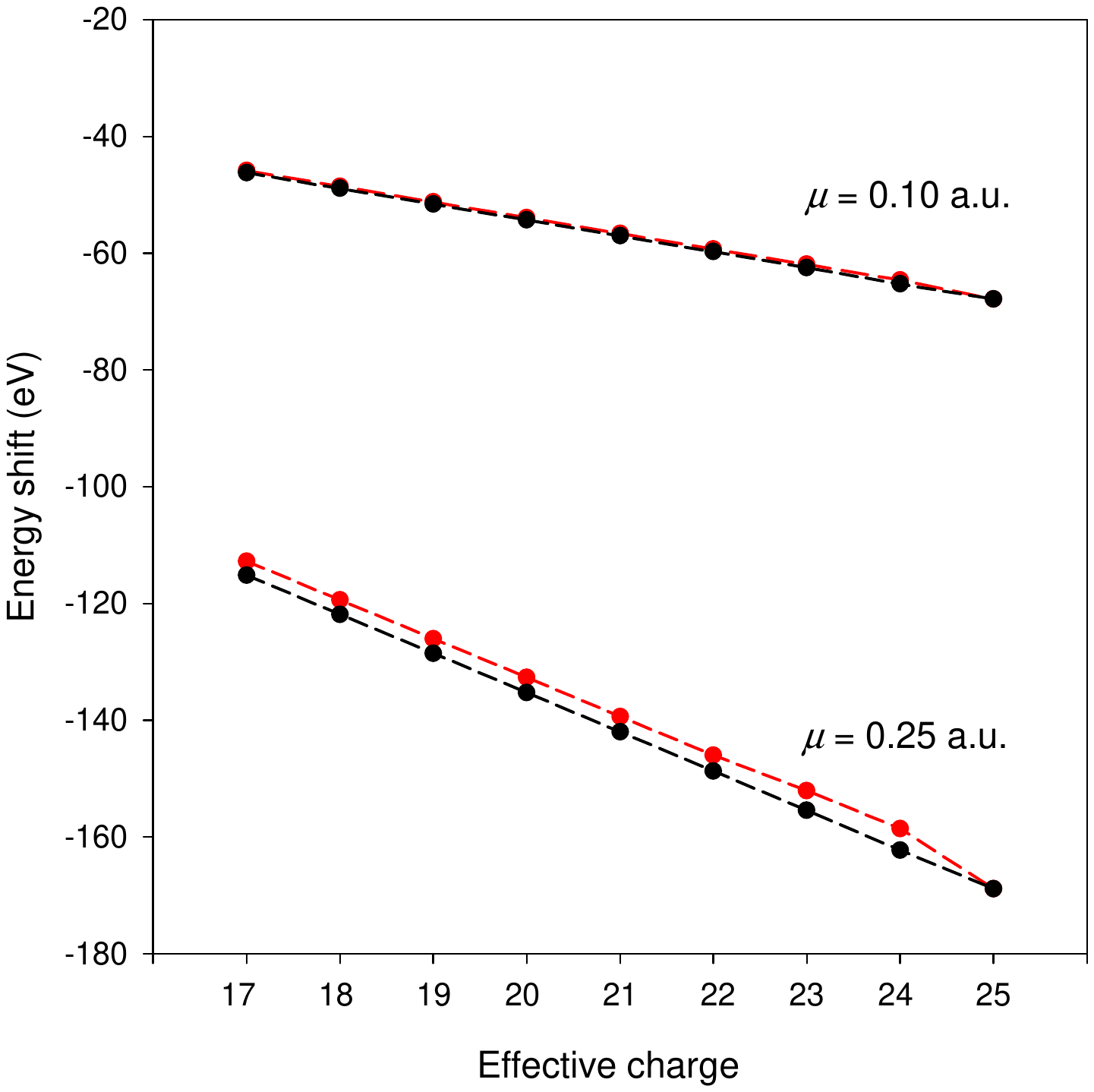}
  \caption{IP (black points) and K-threshold (red points) energy shifts in Fe ions as a function of effective charge $Z_{\rm eff}$ for $\mu = 0.1$~a.u. and $\mu = 0.25$~ a.u.} \label{Fig2}
\end{figure}

This threshold lowering is a well-known phenomenon in dense-plasma physics, whose behavior is further illustrated in Figs.~\ref{Fig1}--\ref{Fig2}. For each of the ionic species considered, the DH potential predicts a linear decrease of the IP downshift with $\mu$ at a gradient that grows with ionic charge (see Fig.~\ref{Fig1}); however, as shown in Fig.~\ref{Fig2}, the IP-shift variation with effective ionic charge $Z_{\rm eff}=Z-N+1$ for a particular $\mu$ is only moderate: ${\sim}30\%$. On the other hand, we find that, for any specific ionic species, the IP and K-threshold energy shifts are practically the same (see Fig.~\ref{Fig2}).

In agreement with \citet{dep19a} and as shown in Table~\ref{IP-plasma-e-e}, the inclusion of the electron--electron Debye screening, in addition to the electron--nucleus screening, leads to a substantially less pronounced IP lowering: 4\% in \ion{Fe}{xxv} to 50\% \ion{Fe}{xvii}. As expected, this discrepancy grows inversely with $Z_{\rm eff}$ as the number of interacting electron--electron pairs is larger for the low-charge states.

The importance of the screening effects by the electron--electron pair interactions has also been recently highlighted by \citet{das16}, who found significant influence of such effects on the IP lowering and excitation energies in Al ions, in particular in the neutral and lowly ionized species. They also found that more stable atomic systems are predicted when the electron--electron screening is taken into account. The Debye electron--nucleus and electron--electron screening effects have been amply discussed by, for instance, \citet{win96}, \citet{kar04, kar05}, \cite{sah06}, \citet{xie12}, \citet{cer13}, \citet{jia14}, and \citet{dep19a}.

\begin{table}[h]
  \caption{Ionization potentials (eV) for \ion{Fe}{xvii} -- \ion{Fe}{xxv} computed with the DH nucleus--electron screening and with/without the electron--electron screening. The parameters $\mu_{ne}$ and $\mu_{ee}$  correspond respectively to the screening parameter $\mu$  (a.u.) used in the first and second terms of Eq.~\eqref{dhpot}. This comparison brings out the importance of the electron--electron screening.}
  \label{IP-plasma-e-e}
  \centering
  \small
  \begin{tabular}{lcccc}
  \hline\hline
  \noalign{\smallskip}
	  &	$\mu_{ne} = 0.1$ & $\mu_{ne} = 0.1$ & $\mu_{ne} = 0.25$ & $\mu_{ne} = 0.25$ \\
  Ion & $\mu_{ee} = 0.0$ & $\mu_{ee} = 0.1$	& $\mu_{ee} = 0.0$  & $\mu_{ee} = 0.25$ \\
  \noalign{\smallskip}
			\hline
  \noalign{\smallskip}
  \ion{Fe}{xvii}  & 1190.72 & 1214.75  & 1089.22 & 1147.80  \\
  \ion{Fe}{xviii} & 1287.21 & 1308.56  & 1185.55 & 1237.68  \\
  \ion{Fe}{xix}   & 1389.10 & 1407.92  & 1287.41 & 1333.09  \\
  \ion{Fe}{xx}    & 1503.54 & 1519.58  & 1401.59 & 1440.80  \\
  \ion{Fe}{xxi}   & 1619.15 & 1632.53  & 1517.02 & 1549.75  \\
  \ion{Fe}{xxii}  & 1727.82 & 1738.54  & 1625.57 & 1651.80  \\
  \ion{Fe}{xxiii} & 1880.58 & 1888.61  & 1778.76 & 1798.43  \\
  \ion{Fe}{xxiv}  & 1974.41 & 1979.79  & 1872.58 & 1885.77  \\
  \ion{Fe}{xxv}   & 8766.28 & 8768.90  & 8661.14 & 8667.86  \\
  \noalign{\smallskip}
  \hline
  \end{tabular}
\end{table}

\subsection{Radiative transitions}

Computed wavelengths and transition probabilities for the stronger K lines ($A>10^{13}$~s$^{-1}$) of \ion{Fe}{xvii} -- \ion{Fe}{xxv} with the three plasma screening parameters  $\mu = 0.0$, 0.1, and 0.25~a.u. are listed in Table~\ref{rad}. If our data for the isolated ion ($\mu$ = 0) are compared with \citet{pal03a} and \citet{men04}, who used the pseudo-relativistic Hartree--Fock (HFR) method, good agreement is generally found. For the transitions listed in Table~\ref{rad}, the wavelengths agree to better than 0.1\% while the transition probabilities can show differences at the 25\% level.  Since we have taken into account similar configuration interaction effects in our atomic models as \citet{pal03a} and \citet{men04}, the main discrepancy source must be attributed to relativistic effects, which are more formerly treated in the MCDF method. Furthermore, we find excellent agreement (within 5\%) among our $A$-values computed in the Babushkin and Coulomb gauges, although only transition probabilities obtained in the Babushkin gauge are reported in Table~\ref{rad}.

\begin{figure}[h!]
  \centering
  \includegraphics[bb=43 259 484 571, clip=true, width=\columnwidth]{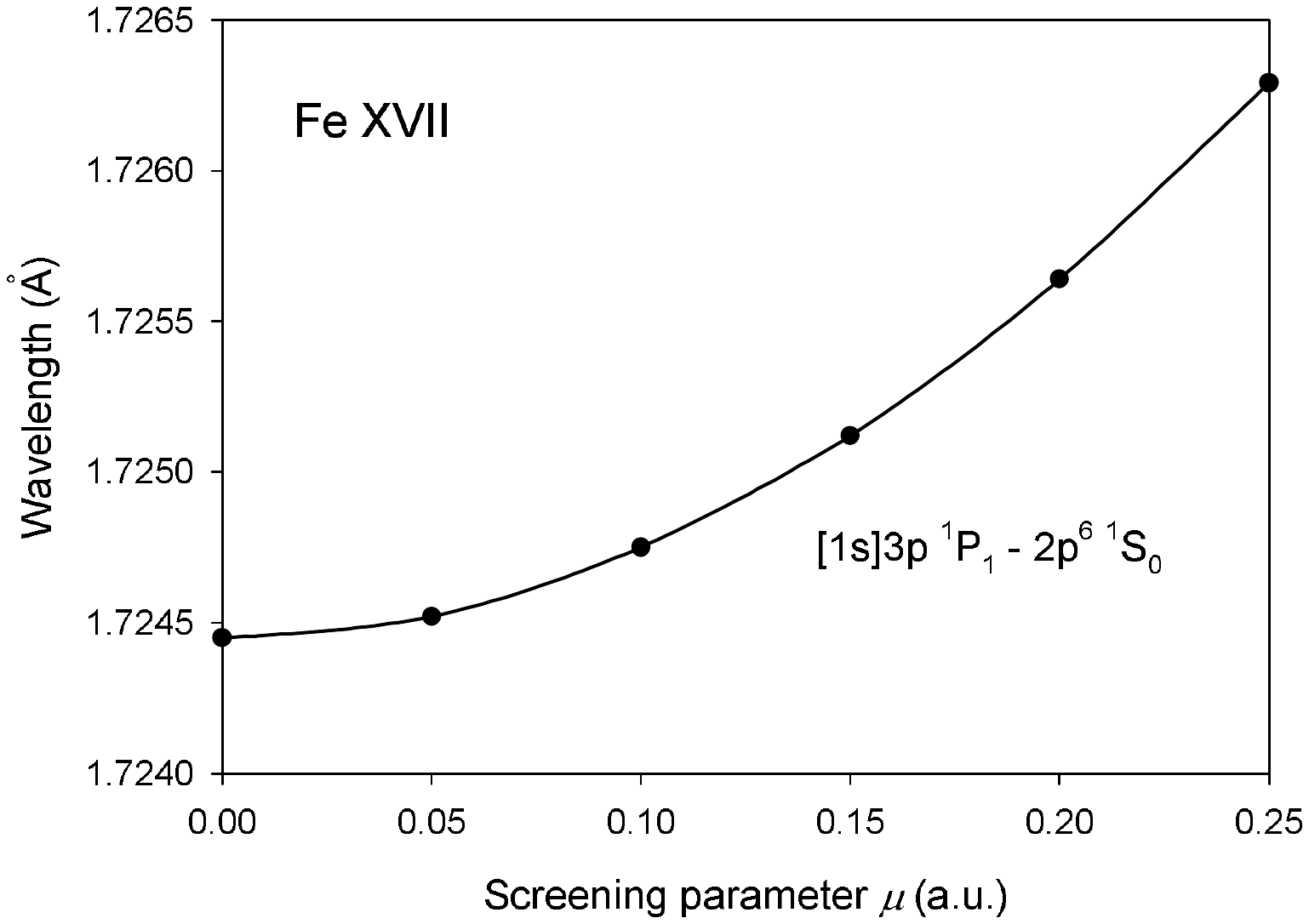}
  \caption{Wavelength reddening as a function of the plasma screening parameter $\mu$ for the [1s]3p $^1$P$_1$ - 2p$^6$ $^1$S$_0$ K$\beta$ line in \ion{Fe}{xvii}.} \label{Fig3}
\end{figure}

\begin{figure}[h!]
  \centering
  \includegraphics[bb=39 196 484 609, clip=true, width=\columnwidth]{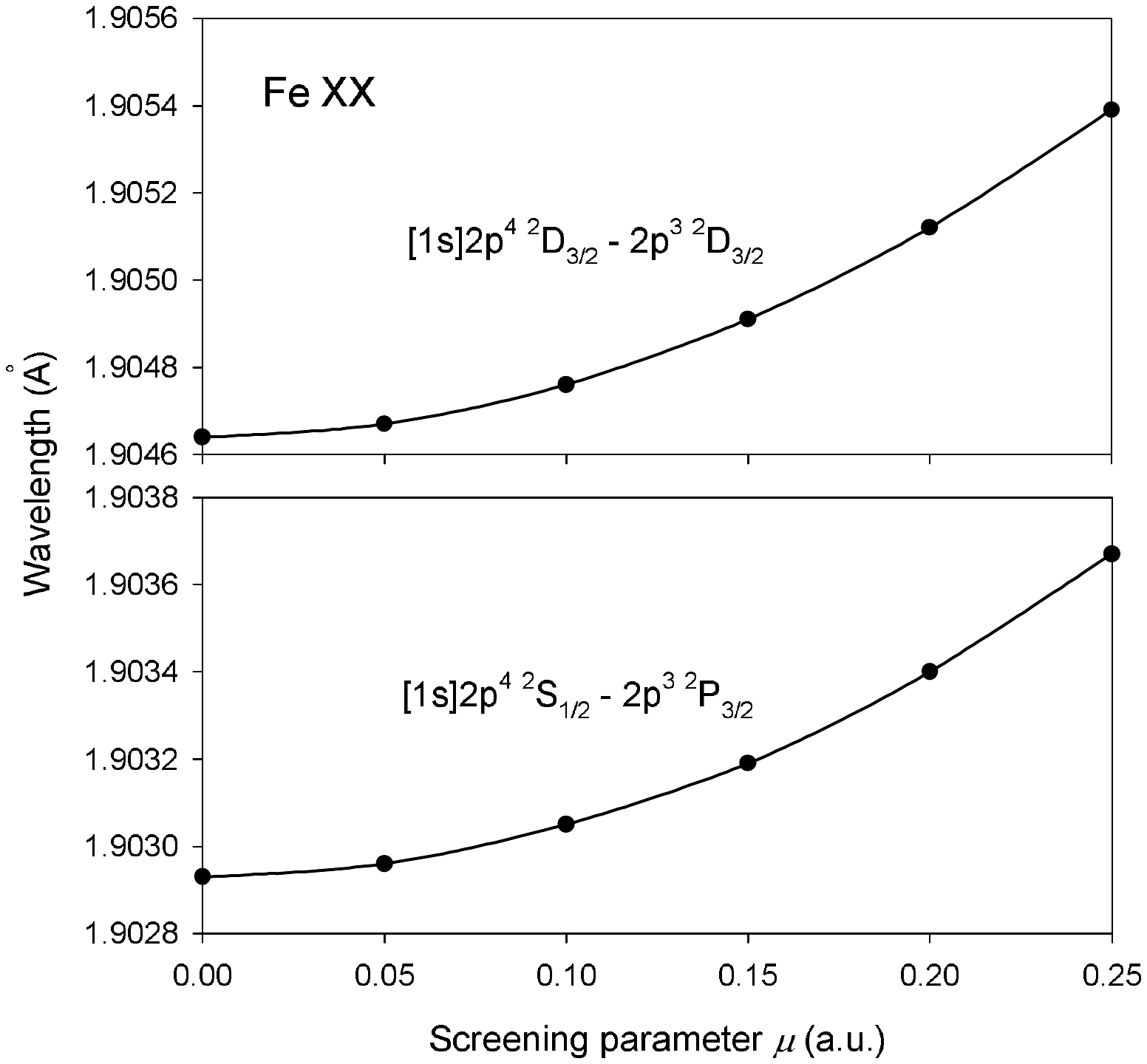}
  \caption{Wavelength reddening as a function of the plasma screening parameter $\mu$ for the [1s]2p$^4$ $^2$D$_{3/2}$ - 2p$^3$ $^2$D$_{3/2}$ and [1s]2p$^4$ $^2$S$_{1/2}$ - 2p$^3$ $^2$P$_{3/2}$ K$_{\alpha}$ lines in \ion{Fe}{xx}.} \label{Fig4}
\end{figure}

A close inspection of Table~\ref{rad} brings out the meager effects of the plasma environment on the K-line radiative parameters. For screening parameters $\mu=0.10$ and $\mu=0.25$, wavelengths appear redshifted relative to those of the isolated ion by less than 0.1~m\AA\ and 1~m\AA, respectively, while the variations of the radiative decay rates do not exceed more than a few percent. To illustrate this point we show in Fig.~\ref{Fig3} the reddening of the [1s]3p $^1$P$_1$ - 2p$^6$ $^1$S$_0$ K$\beta$ line in \ion{Fe}{xvii}, which for $\mu=0.25$ it amounts to ${\sim}2$~m\AA. Similarly, for the [1s]2p$^4$ $^2$D$_{3/2}$ - 2p$^3$ $^2$D$_{3/2}$ and [1s]2p$^4$ $^2$S$_{1/2}$ - 2p$^3$ $^2$P$_{3/2}$ K$\alpha$ lines in \ion{Fe}{xx} in Fig.~\ref{Fig4}, the reddening is less than 1~m\AA.

\begin{figure}[!h]
  \centering
  \includegraphics[bb=77 181 510 597, clip=true, width=\columnwidth]{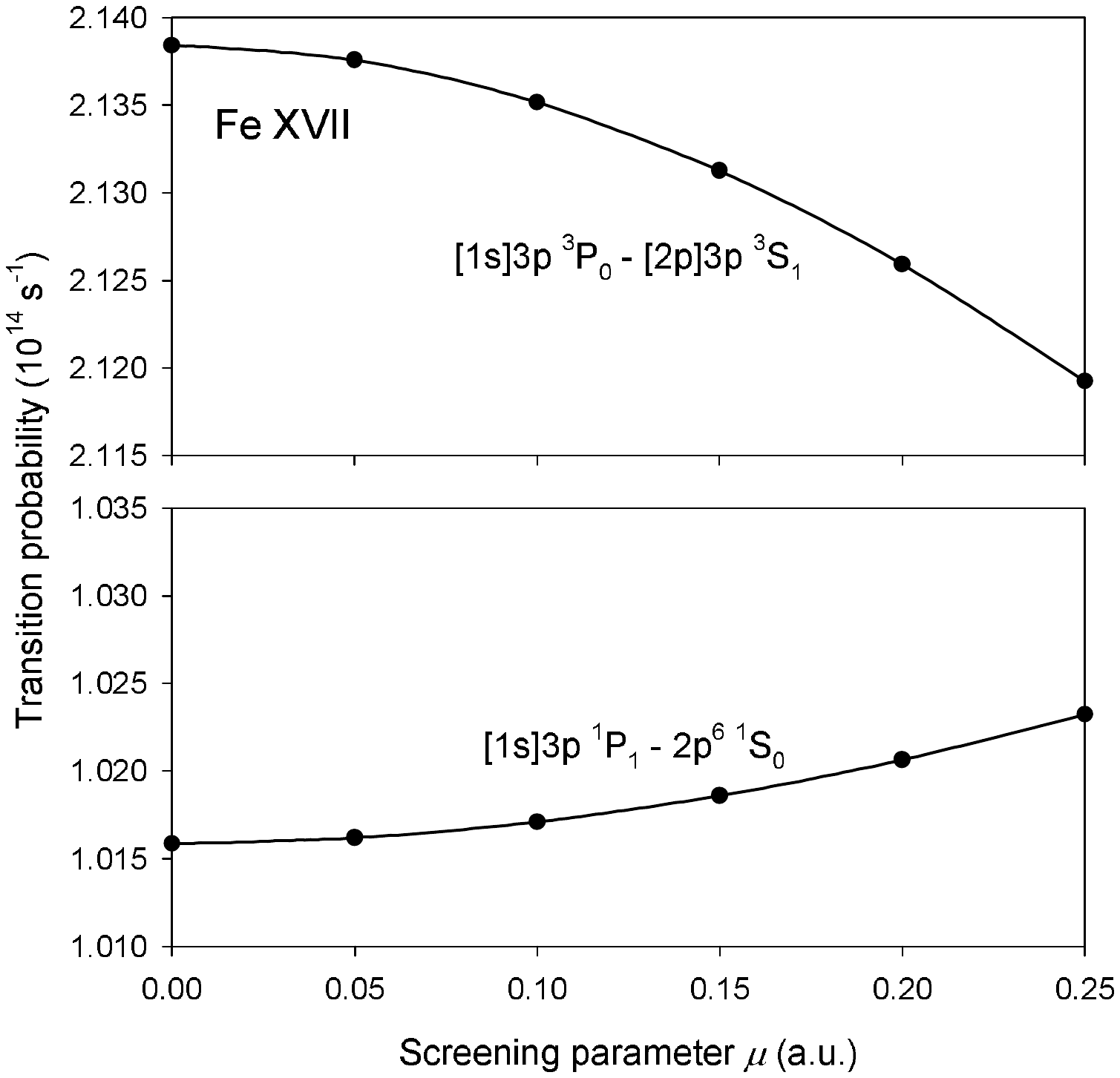}
  \caption{Variation of the radiative transition probability ($A$-value) with the plasma screening parameter $\mu$ for the [1s]3p $^3$P$_0$ - [2p]3p $^3$S$_1$ and [1s]3p $^1$P$_1$ - 2p$^6$ $^1$S$_0$ lines in \ion{Fe}{xvii}.} \label{Fig5}
\end{figure}

\begin{figure}[!h]
  \centering
  \includegraphics[bb=80 174 512 588, clip=true, width=\columnwidth]{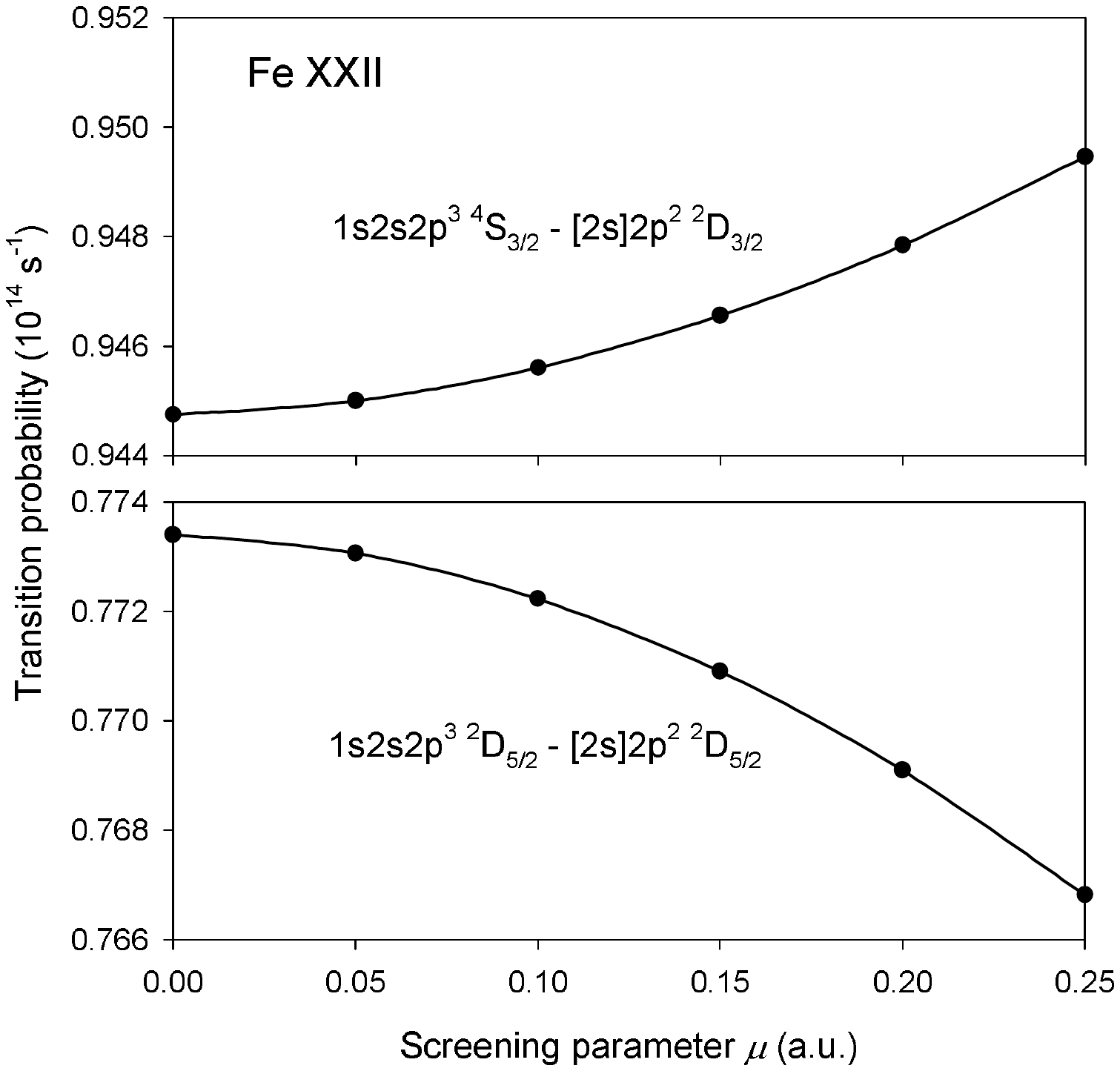}
  \caption{Variation of the transition probability ($A$-value) with the plasma screening parameter $\mu$ for the 1s2s2p$^3$ $^4$S$_{3/2}$ - [2s]2p$^2$ $^2$D$_{3/2}$ and 1s2s2p$^3$ $^2$D$_{5/2}$ - [2s]2p$^2$ $^2$D$_{5/2}$ lines in \ion{Fe}{xxii}.} \label{Fig6}
\end{figure}

It is shown in Fig.~\ref{Fig5} for the [1s]3p $^3$P$_0$ - [2p]3p $^3$S$_1$ and [1s]3p $^1$P$_1$ - 2p$^6$ $^1$S$_0$ lines in \ion{Fe}{xvii} that although the plasma effects on the radiative transition probabilities ($A$-values) are less than 1\%, they can increase or decrease their nominal values ($\mu=0$). In \ion{Fe}{xxii} the changes are somewhat larger (3\%) as illustrated in Fig.~\ref{Fig6} with the 1s2s2p$^3$ $^4$S$_{3/2}$ - [2s]2p$^2$ $^2$D$_{3/2}$ and 1s2s2p$^3$ $^2$D$_{5/2}$ - [2s]2p$^2$ $^2$D$_{5/2}$ lines.

\subsection{Auger widths}

\begin{figure}[!ht]
  \centering
  \includegraphics[bb=80 174 512 588, clip=true, width=\columnwidth]{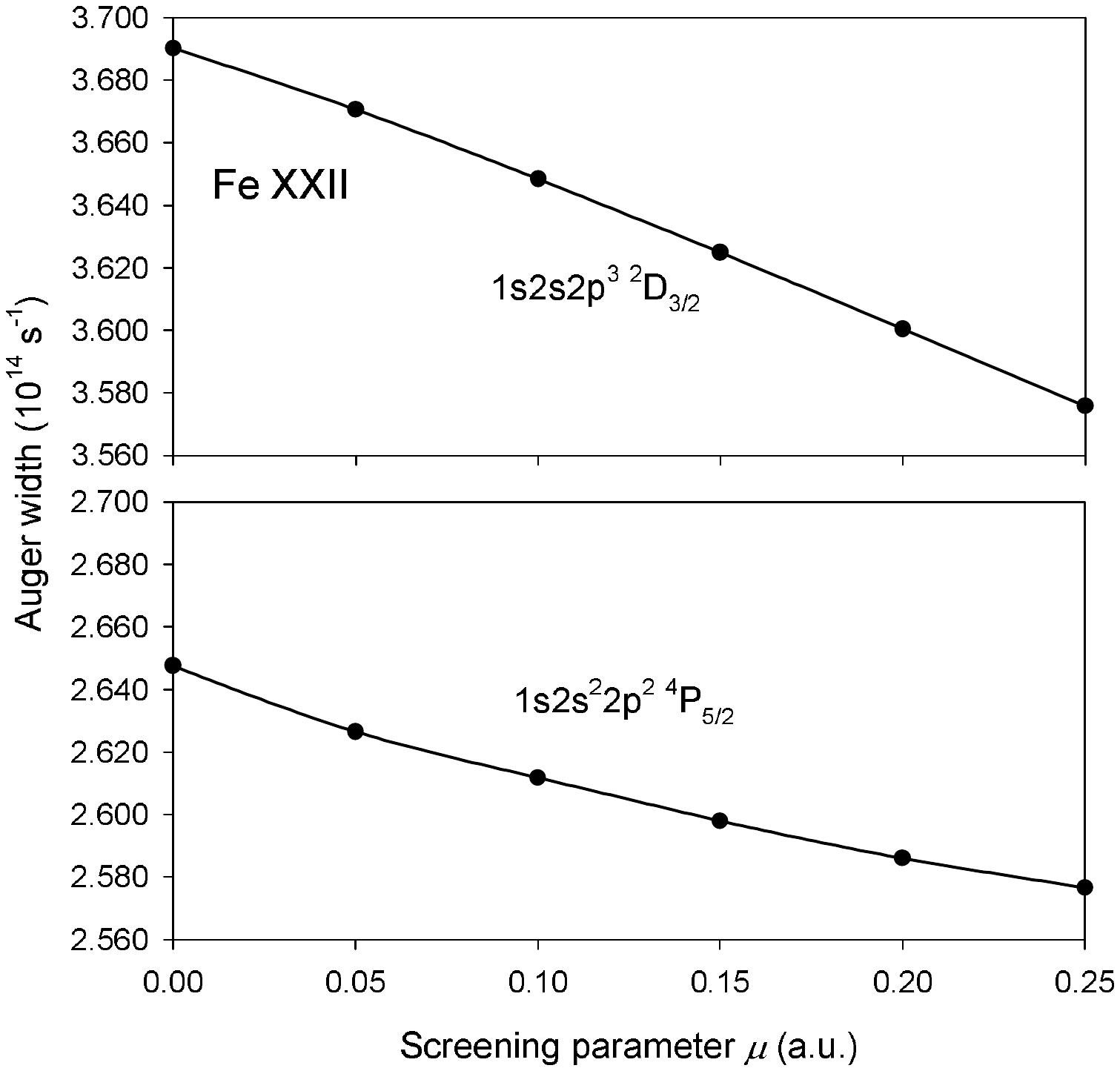}
  \caption{Variation of the Auger width with the plasma screening parameter $\mu$ for two K-vacancy levels in Fe XXII.}
  \label{Fig7}
\end{figure}

Computed Auger widths for the K-vacancy levels with screening parameters $\mu = 0$, 0.1, and 0.25~a.u. are tabulated in Table~\ref{auger}. Present widths for the isolated ion case ($\mu = 0$~a.u.) are found to be in good agreement (within 5\%) with those computed previously with {\sc hfr} by \citet{pal03a}. The DH potential leads to more noticeable decrements of the Auger widths, namely by up to 3\% and 10\% for $\mu = 0.1$ and 0.25~a.u., respectively. This variation is exemplified in Fig.~\ref{Fig7} with the $\mathrm{1s2s2p^3\ ^2D_{3/2}}$  and $\mathrm{1s2s^22p^2\ ^4P_{5/2}}$ K-vacancy levels in \ion{Fe}{xxii}. We also find that the Auger widths for the higher iron ionization stages seem to be more affected by the plasma environment. As previously shown for \ion{Fe}{xvii}, \ion{Fe}{xviii}, and \ion{Fe}{xix} by \citet{dep19b}, due to the weak variations of both the radiative rates and Auger widths with $\mu$, the K-line fluorescence yields in the iron ions considered herein are hardly affected, i.e. by 3\% at most.

\section{Astrophysical implications}

The atomic calculations hereby presented are expected to influence the predictions from plasma modeling in environments where the density and temperature of the gas are such that the screening parameter $\mu$ becomes important, i.e., for relatively low temperatures ($<10^5$~K) and high densities ($> 10^{18}$~cm$^{-3}$, see Table~\ref{Lambda}). These conditions could be encountered in astrophysical environments such as the solar interior (near the convective zone) or in the inner-most regions of accretion disks around compact objects (white dwarfs, neutron stars, and black holes). While the inclusion of the DH potential has a very small effect on the energy levels, radiative probabilities, and Auger widths, it does shift the ionization potential and K-threshold of each ion to lower energies. This effect would modify the part of the ionizing radiation field sampled by the photoionization cross section and, thus, the ionization rates in a photoionized plasma. For simplicity, if we assume a canonical radiation field in the form of an energy power law, the spectrum can be crudely represented by a power law $(h \nu )^{-\alpha}$.  Therefore, for a flat spectrum in energy ($\alpha=1$), the rates will grow proportionally with the reduction of the K-threshold energy leading to a higher ionization rate; moreover, the steeper spectra the larger the change. The emissivity of the fluorescent lines will also display the same dependence increasing the intensity of the observed lines. The heating of the gas due to direct photoionization has the weaker dependence $(h\nu)^{1-\alpha}$.

The shifts in the K thresholds are of ${\sim}110{-}170$~eV for \ion{Fe}{xvii} -- \ion{Fe}{xxv} at the largest screening parameter considered here. These shifts are marginally too small to be detected with the available X-ray observatories. Nevertheless, changes in the Fe K-edge position will be resolved with new instruments such as the micro-calorimeters aboard the future missions {\it XRISM} \citep{tas18,gua18} and {\it ATHENA} \citep{nan13,gua18}, which will have a resolution of a few eVs. Notice that unlike the energies of the K-thresholds, the energies of the K$\alpha$ and K$\beta$ lines are almost unaffected by high density plasma effects. Hence, high spectral resolution observations should yield information on the ionization structure of the plasma and plasma effect simultaneously. This is by constraining the presence of different ionic stages on the position and structure of the K-lines, while diagnosing plasma screening effects on the position and structure of the respective K-thresholds. Moreover, we emphasize that the effects mentioned here only concern changes in the ionic radiative properties. Other plasma effects at high densities, such as the continuum lowering and increase of the collisional rates, are expected to introduce additional modifications to the observed spectra. These effects will be a matter of further study in our on-going projects.

\section{Summary and conclusion}

The influence of the plasma environment on the atomic structure and the radiative and Auger properties of K lines in highly charged iron ions (from He-like \ion{Fe}{xxv} to Ne-like \ion{Fe}{xvii}) has been studied by means of a time-averaged Debye--H\"uckel screened potential in the context of the relativistic multiconfiguration Dirac--Fock framework. The explored plasma screening-parameter space of the problem spans $0\leq\mu\leq 0.25$~a.u. corresponding to conditions expected in compact-object accretion disks. The results can be summarized as follows:
\begin{itemize}
\item[1.] The ionization potentials and K-threshold energies are both lowered by the plasma environment. The lowering is found to be linearly dependent on both the plasma screening parameter $\mu$ and the ionic effective charge $Z_{\rm eff}$. They vary from ${\sim}110$~eV to ${\sim}170$~eV for $\mu = 0.25~{\rm a.u.}$ corresponding to extreme plasma conditions.
\item[2.] The importance of the electron--electron plasma screening has been reconfirmed.
\item[3.] Plasma environment effects are negligibly small on the wavelengths (redshifted by less than $\sim 0.1~\rm{\AA}$), the transition probabilities (they decrease or increase depending on transitions by a few percent) and the Auger widths (they decrease by at most $\sim$10\%).

\end{itemize}

In conclusion, the high-resolution X-ray spectrometers onboard future missions such as {\it XRISM} and {\it ATHENA} are expected to be sensitive to the lowering of the iron K edge due to the extreme plasma conditions occurring in accretion disks around compact objects.

\begin{acknowledgements}
J.D. is Research Fellow of the Belgian Fund for Research Training in Industry and Agriculture (FRIA) while P.P. and P.Q. are, respectively, Research Associate and Research Director of the Belgian Fund for Scientific Research (F.R.S.-FNRS). Financial supports from these organizations, as well as from the NASA Astrophysics Research and Analysis Program (grant 80NSSC17K0345) are gratefully acknowledged. J.A.G. acknowledges support from the Alexander von Humboldt Foundation.
\end{acknowledgements}

\begin{table*}[t!]
  \caption{Comparison of wavelengths and transition probabilities for the K lines of \ion{Fe}{xvii} -- \ion{Fe}{xxv} ($17\leq Z_{\rm eff}\leq 25$) computed with plasma screening parameters $\mu = 0.0$, 0.1, and 0.25~a.u. Data obtained with $\mu = 0$~a.u. corresponds to an isolated ion. \label{rad}}
  \centering
  \small
  \begin{tabular}{clcccccc}
  \hline\hline
  \noalign{\smallskip}
  $Z_{\rm eff}$ & Transition & \multicolumn{3}{c}{Wavelength (\AA)} &  \multicolumn{3}{c}{Transition probability (s$^{-1}$)} \\
  & & $\mu$ = 0.0 & $\mu$ = 0.1 & $\mu$ = 0.25  & $\mu$ = 0.0 & $\mu$ = 0.1 & $\mu$ = 0.25  \\
  \hline
  \noalign{\smallskip}
  17 & $\mathrm{[1s]3p\ ^1P_1-2p^6\ ^1S_0}$   &	1.7244 & 1.7248	& 1.7263 & 1.016E+14 & 1.017E+14 & 1.023E+14 \\
  17 & $\mathrm{[1s]3p\ ^1P_1-[2p]3p\ ^3S_1}$ &	1.9253 & 1.9254	& 1.9260 & 1.131E+13 & 1.130E+13 & 1.127E+13 \\
  17 & $\mathrm{[1s]3p\ ^3P_2-[2p]3p\ ^3S_1}$ &	1.9263 & 1.9264	& 1.9270 & 2.370E+13 & 2.373E+13 & 2.386E+13 \\
  17 & $\mathrm{[1s]3p\ ^1P_1-[2p]3p\ ^3D_2}$ &	1.9265 & 1.9267	& 1.9273 & 2.919E+13 & 2.927E+13 & 2.964E+13 \\
  17 & $\mathrm{[1s]3s\ ^1S_0-[2p]3s\ ^1P_1}$ &	1.9268 & 1.9269	& 1.9275 & 3.225E+14 & 3.223E+14 & 3.211E+14 \\
  17 & $\mathrm{[1s]3p\ ^3P_1-[2p]3p\ ^3S_1}$ &	1.9270 & 1.9271	& 1.9277 & 8.108E+13 & 8.109E+13 & 8.112E+13 \\
  17 & $\mathrm{[1s]3p\ ^3P_0-[2p]3p\ ^3S_1}$ &	1.9272 & 1.9273	& 1.9279 & 2.138E+14 & 2.135E+14 & 2.119E+14 \\
  17 & $\mathrm{[1s]3p\ ^1P_1-[2p]3p\ ^1P_1}$ &	1.9274 & 1.9275	& 1.9281 & 1.108E+14 & 1.107E+14 & 1.103E+14 \\
  17 & $\mathrm{[1s]3s\ ^3S_1-[2p]3s\ ^3P_2}$ &	1.9278 & 1.9279	& 1.9285 & 3.401E+14 & 3.400E+14 & 3.394E+14 \\
  17 & $\mathrm{[1s]3p\ ^3P_2-[2p]3p\ ^3D_3}$ &	1.9280 & 1.9281	& 1.9287 & 2.871E+14 & 2.870E+14 & 2.865E+14 \\
  17 & $\mathrm{[1s]3p\ ^1P_1-[2p]3p\ ^3P_2}$ &	1.9280 & 1.9281	& 1.9287 & 1.505E+14 & 1.504E+14 & 1.497E+14 \\
  17 & $\mathrm{[1s]3p\ ^3P_1-[2p]3p\ ^3D_2}$ &	1.9283 & 1.9284	& 1.9290 & 3.090E+14 & 3.088E+14 & 3.079E+14 \\
  17 & $\mathrm{[1s]3p\ ^3P_2-[2p]3p\ ^1P_1}$ &	1.9284 & 1.9285	& 1.9291 & 1.453E+13 & 1.450E+13 & 1.435E+13 \\
  17 & $\mathrm{[1s]3s\ ^3S_1-[2p]3s\ ^1P_1}$ &	1.9285 & 1.9286	& 1.9292 & 9.717E+13 & 9.718E+13 & 9.724E+13 \\
  17 & $\mathrm{[1s]3p\ ^3P_2-[2p]3p\ ^3P_2}$ &	1.9290 & 1.9291	& 1.9297 & 9.841E+13 & 9.840E+13 & 9.834E+13 \\
  17 & $\mathrm{[1s]3p\ ^3P_0-[2p]3p\ ^1P_1}$ &	1.9293 & 1.9294	& 1.9300 & 2.092E+14 & 2.094E+14 & 2.102E+14 \\
  17 & $\mathrm{[1s]3p\ ^3P_1-[2p]3p\ ^3P_2}$ &	1.9297 & 1.9298	& 1.9304 & 2.434E+13 & 2.439E+13 & 2.461E+13 \\
  17 & $\mathrm{[1s]3p\ ^1P_1-[2p]3p\ ^3P_0}$ &	1.9297 & 1.9298	& 1.9305 & 1.148E+13 & 1.143E+13 & 1.119E+13 \\
  17 & $\mathrm{[1s]3p\ ^1P_1-[2p]3p\ ^3D_1}$ &	1.9302 & 1.9303	& 1.9309 & 3.116E+13 & 3.121E+13 & 3.145E+13 \\
  17 & $\mathrm{[1s]3s\ ^1S_0-[2p]3s\ ^3P_1}$ &	1.9304 & 1.9305	& 1.9311 & 2.896E+14 & 2.896E+14 & 2.899E+14 \\
  17 & $\mathrm{[1s]3p\ ^1P_1-[2p]3p\ ^3P_1}$ &	1.9312 & 1.9313	& 1.9319 & 5.188E+13 & 5.184E+13 & 5.167E+13 \\
  17 & $\mathrm{[1s]3p\ ^1P_1-[2p]3p\ ^3D_2}$ &	1.9313 & 1.9314	& 1.9320 & 1.580E+14 & 1.580E+14 & 1.577E+14 \\
  17 & $\mathrm{[1s]3p\ ^3P_1-[2p]3p\ ^3P_0}$ &	1.9315 & 1.9316	& 1.9322 & 5.507E+13 & 5.510E+13 & 5.523E+13 \\
  17 & $\mathrm{[1s]3s\ ^3S_1-[2p]3s\ ^3P_0}$ &	1.9316 & 1.9317	& 1.9324 & 6.669E+13 & 6.667E+13 & 6.656E+13 \\
  17 & $\mathrm{[1s]3p\ ^3P_1-[2p]3p\ ^3D_1}$ &	1.9319 & 1.9320	& 1.9327 & 1.099E+14 & 1.098E+14 & 1.093E+14 \\
  17 & $\mathrm{[1s]3s\ ^3S_1-[2p]3s\ ^3P_1}$ &	1.9321 & 1.9322	& 1.9328 & 1.042E+14 & 1.042E+14 & 1.038E+14 \\
  17 & $\mathrm{[1s]3p\ ^3P_0-[2p]3p\ ^3D_1}$ &	1.9321 & 1.9322	& 1.9328 & 1.694E+14 & 1.693E+14 & 1.689E+14 \\
  17 & $\mathrm{[1s]3p\ ^3P_2-[2p]3p\ ^3P_1}$ &	1.9322 & 1.9323	& 1.9329 & 7.919E+13 & 7.915E+13 & 7.896E+13 \\
  17 & $\mathrm{[1s]3p\ ^3P_2-[2p]3p\ ^3D_2}$ &	1.9323 & 1.9324	& 1.9331 & 1.014E+14 & 1.013E+14 & 1.010E+14 \\
  17 & $\mathrm{[1s]3p\ ^3P_0-[2p]3p\ ^3P_1}$ &	1.9331 & 1.9332	& 1.9338 & 1.615E+13 & 1.619E+13 & 1.637E+13 \\
  17 & $\mathrm{[1s]3p\ ^1P_1-[2p]3p\ ^1S_0}$ &	1.9380 & 1.9381	& 1.9388 & 4.421E+13 & 4.421E+13 & 4.421E+13 \\
  17 & $\mathrm{[1s]3p\ ^3P_1-[2p]3p\ ^1S_0}$ &	1.9398 & 1.9399	& 1.9406 & 1.056E+13 & 1.051E+13 & 1.029E+13 \\
  18 & $\mathrm{[1s]2p^6\ ^2S_{1/2}-2p^5\ ^2P_{3/2}}$  & 1.9262	& 1.9263 & 1.9268    & 4.116E+14 & 4.115E+14 & 4.103E+14 \\
  18 & $\mathrm{[1s]2p^6\ ^2S_{1/2}-2p^5\ ^2P_{1/2}}$  & 1.9301	& 1.9302 & 1.9307    & 2.018E+14 & 2.017E+14 & 2.012E+14 \\
  \hline
  \end{tabular}
  \tablefoot{A complete version of this table is available in electronic form from the CDS.}
\end{table*}

\begin{table*}[h!]
  \caption{Plasma environment effects on the Auger widths of K-vacancy states in \ion{Fe}{xvii} -- \ion{Fe}{xxv} ($17\leq Z_{\rm eff}\leq 25$) computed with plasma screening parameters $\mu = 0.0$, 0.1, and 0.25~a.u. Data obtained with $\mu = 0$~a.u. corresponds to an isolated ion. \label{auger}}
  \centering
  \small
  \begin{tabular}{clccc}
  \hline\hline
  \noalign{\smallskip}
  $Z_{\rm eff}$ & Level & \multicolumn{3}{c}{Auger width (s$^{-1}$)}  \\
  &	    & $\mu = 0.0$ & $\mu = 0.1$  &  $\mu = 0.25$ \\
  \hline
  \noalign{\smallskip}	
  17 & $\mathrm{[1s]3s\ ^3S_1}$       & 7.711E+14 & 7.647E+14 & 7.604E+14 \\
  17 & $\mathrm{[1s]3s\ ^1S_0}$       & 8.135E+14 & 8.069E+14 & 8.027E+14 \\
  17 & $\mathrm{[1s]3p\ ^3P_0}$       & 7.610E+14 & 7.548E+14 & 7.420E+14 \\
  17 & $\mathrm{[1s]3p\ ^3P_1}$       & 7.302E+14 & 7.243E+14 & 7.128E+14 \\
  17 & $\mathrm{[1s]3p\ ^3P_2}$       & 7.154E+14 & 7.097E+14 & 7.069E+14 \\
  17 & $\mathrm{[1s]3p\ ^1P_1}$       & 7.244E+14 & 7.189E+14 & 7.106E+14 \\
  18 & $\mathrm{[1s]2p^6\ ^2S_{1/2}}$ & 1.529E+15 & 1.523E+15 & 1.504E+15 \\
  19 & $\mathrm{[1s]2p^5\ ^3P_2}$     & 8.058E+14 & 8.016E+14 & 7.951E+14 \\
  19 & $\mathrm{[1s]2p^5\ ^3P_1}$     & 7.929e+14 & 7.890E+14 & 7.828E+14 \\
  19 & $\mathrm{[1s]2p^5\ ^3P_0}$     & 7.803E+14 & 7.764E+14 & 7.699E+14 \\
  19 & $\mathrm{[1s]2p^5\ ^1P_1}$     & 7.512E+14 & 7.484E+14 & 7.427E+14 \\
  19 & $\mathrm{1s2s2p^6\ ^3S_1}$     & 7.663E+14 & 7.623E+14 & 7.598E+14 \\
  19 & $\mathrm{1s2s2p^6\ ^1S_0}$     & 1.137E+15 & 1.130E+15 & 1.115E+15 \\
  \hline	
  \end{tabular}
  \tablefoot{A complete version of this table is available in electronic form from the CDS.}
\end{table*}

\end{document}